\documentstyle[12pt,fleqn]{article}
\def\singlespace {\smallskipamount=3.75pt plus1pt minus1pt
                  \medskipamount=7.5pt plus2pt minus2pt
                  \bigskipamount=15pt plus4pt minus4pt
                  \normalbaselineskip=15pt plus0pt minus0pt
                  \normallineskip=1pt
                  \normallineskiplimit=0pt
                  \jot=3.75pt
                  {\def\smallskip {\vskip\smallskipamount}}
                  {\def\medskip   {\vskip\medskipamount}}
                  {\def\bigskip   {\vskip\bigskipamount}}
                  {\setbox\strutbox=\hbox{\vrule 
                    height10.5pt depth4.5pt width 0pt}}
                  \parskip 7.5pt
                  \normalbaselines}
\def\middlespace {\smallskipamount=5.625pt plus1.5pt minus1.5pt
                  \medskipamount=11.25pt plus3pt minus3pt
                  \bigskipamount=22.5pt plus6pt minus6pt
                  \normalbaselineskip=22.5pt plus0pt minus0pt
                  \normallineskip=1pt
                  \normallineskiplimit=0pt
                  \jot=5.625pt
                  {\def\smallskip {\vskip\smallskipamount}}
                  {\def\medskip   {\vskip\medskipamount}}
                  {\def\bigskip   {\vskip\bigskipamount}}
                  {\setbox\strutbox=\hbox{\vrule 
                    height15.75pt depth6.75pt width 0pt}}
                  \parskip 11.25pt
                  \normalbaselines}
\def\doublespace {\smallskipamount=7.5pt plus2pt minus2pt
                  \medskipamount=15pt plus4pt minus4pt
                  \bigskipamount=30pt plus8pt minus8pt
                  \normalbaselineskip=30pt plus0pt minus0pt
                  \normallineskip=2pt
                  \normallineskiplimit=0pt
                  \jot=7.5pt
                  {\def\smallskip {\vskip\smallskipamount}}
                  {\def\medskip   {\vskip\medskipamount}}
                  {\def\bigskip   {\vskip\bigskipamount}}
                  {\setbox\strutbox=\hbox{\vrule 
                    height21.0pt depth9.0pt width 0pt}}
                  \parskip 15.0pt
                  \normalbaselines}

\include{dspace12}
\def\be{\begin{equation}}
\def\ee{\end{equation}}
\def\bea{\begin{eqnarray}}
\def\eea{\end{eqnarray}}
\def\nn{\nonumber}
\def\th{\theta}

\def\lt{\left}
\def\rt{\right}
\def\sect #1{\setcounter{equation}{0}}

\begin{document}
\singlespace

\vspace{0.5in}

\begin{center}
{\LARGE { Directional naked singularity in general relativity
}}
\end{center}
\vspace{0.7in}
\vspace{12pt}
\begin{center}
{\large{ 
K. S. Virbhadra\footnotemark[1]\\
Theoretical Astrophysics Group\\
Tata Institute of Fundamental Research\\
Homi Bhabha Road, Colaba, Bombay 400005, India.\\
}}
\end{center}
\footnotetext[1]{E-mail : shwetketu@tifrvax.tifr.res.in}
\vspace{1.3in}
\begin{abstract}
We consider a static, axially symmetric, and asymptotically flat
exact solution of the Einstein vacuum equations, known as the
gamma metric. This is characterized by two constant parameters
$m$ and $\gamma$. We find that the total energy associated with
this metric is $m \gamma$. Considering the total energy to be
positive, we investigate the nature of a curvature singularity $r=2m$
($r$ is the radial coordinate) in this metric. 
For $\gamma < 1$, this singularity is globally visible along $\theta = 0$
as well as $\theta = \pi /2$. However, for $\gamma > 1$, this singularity is
though globally naked along $\theta =\pi/2$, it is not visible
(even locally) along $\theta = 0$. Thus, this  exhibits
``directional nakedness'' for $\gamma > 1$. This could have implications
for astrophysics.
\end{abstract}
\newpage
It is well-known that the general theory of relativity predicts
the occurrence of the spacetime singularities in gravitational
collapse.  At singularities, the spacetime curvature is enormously
large and the classical general relativity theory breaks down there.
Whether or not the singularities are visible to observers at infinity has been
debated. As physics at a spacetime singularity is not known, the
existence of a naked singularity is usually expected to give serious
problems as compared to a singularity which is not visible. For
instance, there can be production of matter and/or radiation out
of extremely high gravitational field and, as one knows, mechanism for that is
not understood. Due to such reasons, naked singularities are abhorant to many physicists.
The problem is observationally avoided if and only if it is assumed
that no information can escape out of a singularity. Penrose\cite{Pen69}, 
in a seminal
review, asked, ``Does there exist a cosmic censor who forbids the
occurrence of naked singularities, clothing each one in an
absolute event horizon?''
The answer to this question is not known. Penrose\cite{Pen78} as well as
Hawking\cite{Haw79} considered this as the most important unsolved
problem of classical general relativity theory. The hypothesis
that a physically realistic collapse will not lead to naked
singularities is referred to as the cosmic censorship hypothesis
(CCH)(\cite{Pen69}-\cite{Pen78}). Penrose\cite{Pen78} mentioned that unless the
production of a naked singularity in a gravitational collapse is stable,
the CCH remains valid. There is no agreed and precise statement of a
CCH. There exists in the literature some other formulations to 
CCH\cite{PenSeiNew}. However, due to the lack of
a precise mathematical formulation describing ``a physically
realistic system'', no proof for any version of CCH is known.
 
Penrose\cite{Pen72}, three years after he proposed the CCH, expressed his
opinion that it had long seemed to him that the presumption that the spacetime 
singularities that arise in gravitational collapse must inevitably reside
inside black holes was a product largely of  wishful thinking. He further
suggested that the possibility that naked singularities may sometimes arise
must be considered seriously. Since many remarkable and violent phenomena are 
seen in astronomy, he\cite{Pen73}  suggested that it is worth speculating that 
naked singularities may  play a role. Whether or not  the  CCH is true is a
very important issue, as its validity is fundamental to lot of work that has
been done on black holes.  On the other hand, if the CCH 
is wrong and naked singularities occur in nature, then one might have a
chance to study the effects of highly curved regions of spacetime.
Over last twenty five years, after the CCH was proposed, the
subject of singularity has fascinated many researchers' minds(\cite{
HawWalNabCla}-\cite{UnnAnt}). When a  proof for a hypothesis is not known,
it is worth obtaining counterexamples (see \cite{NS}-\cite{NSP} and
 references therein).

The Schwarzschild solution has a spacetime singularity at $r=0$
($r$ is the radial coordinate). This singularity is covered by an
event horizon if the mass parameter $m$ is positive. Similarly,
the Kerr solution has a ring singularity which is covered by an
event horizon if the mass parameter $m$ is positive and is greater
than the absolute value of the rotation parameter $a$. However, for
the case $m^2<a^2$, there is no event horizon and the singularity
is globally naked. Carter demonstrated that the ring singularity 
($m^2<a^2$) is visible only for the equatorial plane (see in
\cite{Pen73}). He also showed that the null
geodesics from the neighbourhood of singular ring can escape to infinity
only in directions very close to the equatorial plane.
Using this property, Penrose (\cite{Pen72},\cite{Pen73}) suggested a possible
explanation to Weber's gravitational waves observation. 
He argued that a rotating naked singularity at the center of the
Galaxy would have the property that signals from the
neighbourhood of this are necessarily beamed in one plane. This could remove
the mass-loss conflict in Weber's observation. 
However, it is now usually believed (though it has not been explicitly 
proved) that $m^2<a^2$ Kerr  singularity cannot result from a realistic 
gravitational collapse. Though there is a general
consensus that Weber's conclusion was wrong, it remains of interest
to investigate whether or not other visible singularities have similar
characteristics, which could be useful for explaining some astronomical
observations in future.  Obviously, only aspherical visible singularities
can show such behaviour.  We investigate the well-known gamma
metric and find that it has a singularity which possesses ``directional
nakedness''. We use geometrized units ( $G =1 , c =1$)  and follow the
convention that Latin (Greek)  indices take values $0\ldots3$ ($1\ldots3$).

A static and asymptotically flat exact
solution to the Einstein vacuum equations, known as the gamma
metric, is given by the line element\cite{Ste}:
\be
ds^2 =\lt(1 - \frac{2m}{r}\rt)^{\gamma} dt^2 
     - \lt(1 - \frac{2m}{r}\rt)^{-\gamma} 
      \lt[\lt(\frac{\Delta}{\Sigma}\rt)^{\gamma^2-1} dr^2 
     + \frac{\Delta^{r^2}}{\Sigma^{\gamma^2-1}} d\theta^2 
      + \Delta \sin^2\theta  d\phi^2\rt] ,
\label{eq1}
\ee
where,
\bea
\Delta &=& r^2 - 2mr, \nn\\
\Sigma &=& r^2 -2mr + m^2 \sin^2 \theta .
\label{eq2}
\eea
$m$ and $\gamma$ are two constant parameters in this solution.
$m=0$ or $\gamma=0$ gives the flat spacetime. For $|\gamma|=1$
the metric is spherically symmetric and for $|\gamma| \neq 1$,
it is axially symmetric.  $\gamma=1$ gives the Schwarzschild 
spacetime in the Schwarzschild coordinates.
$\gamma =-1$ gives the Schwarzschild spacetime with negative
mass, as putting $m=-M (M>0)$ and carrying out a nonsingular
coordinate transformation ($r \rightarrow R = r + 2 M$) one gets
the  Schwarzschild spacetime (with positive mass) in the Schwarzschild 
coordinates $t,R,\theta,\Phi$.

First, we investigate the total energy, momentum, and angular
momentum for the gamma metric. For this purpose, we use the
symmetric energy-momentum complex of Weinberg\cite{Wei}, which is given
by
\begin{equation}
W^{ik}= \frac{1}{16 \pi} {D^{lik}}_{,l},
\label{eq3}
 \end{equation}
where
 \begin{equation}
D^{lik}= \frac{\partial h^a_ a}{\partial x_l}\eta^{ik}
         -  \frac{\partial h^a_ a}{\partial x_i}\eta^{lk}
         - \frac{\partial h^{al}}{\partial x^a}\eta^{ik}
         + \frac{\partial h^{ai}}{\partial x^a}\eta^{lk}
         + \frac{\partial h^{lk}}{\partial x_i}
         - \frac{\partial h^{ik}}{\partial x_l}
\label{eq4}
 \end{equation}
with
 \begin{equation}
h_{ik}=g_{ik} - \eta_{ik}.
\label{eq5}
 \end{equation}
$\eta_{ik}$ is the Minkowski metric. Indices on $h_{ik}$ or $\partial/
\partial x_i$ are raised or lowered with help of $\eta$'s.
It is obvious that
 \begin{equation}
D^{lik}=- D^{ilk}.
\label{eq6}
 \end{equation}
$W^{00}$ and $W^{\alpha 0}$  are the energy and energy current (momentum) 
density components, respectively. $W^{ik}$ satisfies the  covariant local
conservation laws:
 \begin{equation}
\frac{\partial W^{ik}}{\partial  x^k}=0.
\label{eq7}
 \end{equation}
Using Gauss's theorem, one has the energy and momentum components
($P^0$ is the energy and $P^\alpha$ are the momentum components)
 \begin{equation}
P^i=\frac{1}{16 \pi} \int\!\!\!\int{ D^{\alpha 0 i} n_{\alpha}
\, dS}
\label{eq8}
 \end{equation}
and the physically interesting components of the angular momentum are
 \begin{equation}
J^{\alpha \beta}=\frac{1}{16 \pi} \int\!\!\!\int{
\left(x^{\alpha} D^{\sigma 0 \beta}
-x^{\beta} D^{\sigma 0 \alpha}
+\eta^{\sigma\alpha} h^{0\beta}
- \eta^{\sigma\beta} h^{0\alpha}
\right)
 n_{\sigma}\, dS}.
\label{eq9}
 \end{equation}
where $n_{\alpha}$ is the outward unit normal vector and $dS$ is the
infinitesimal surface element.

The use of the Weinberg energy-momentum complex (which is a non-tensorial
object), like any other energy-momentum complex, is restricted to
quasi-Minkowskian coordinates (see \cite{ACV} and references therein).
Therefore, we transform the line element $(\ref{eq1})$ to quasi-Minkowskian
coordinates ($t,x,y,z$)  according to $ x = r \sin\theta \cos\phi,
y = r \sin\theta\sin\phi$ and $ z = r \cos\theta$, and evaluate
the above integrals at large distance. We get the total  energy,
momentum, and angular momentum :
\bea
P^0 &=& m \gamma , \nn\\
P^{\alpha} &=& 0, \nn\\
J^{\alpha\beta} &=& 0.
\label{eq10}
\eea
Respecting the total energy to be nonnegative, in the following,
we consider $m>0,\gamma>0$, but $\gamma \neq 1$ (as $\gamma = 1$
corresponds to the Schwarzschild metric and the structure of
singularity for that is well-known in the literature). The
divergence of the Kretschmann invariant ${\cal{K}} \equiv R_{abcd}R^{abcd}$
($R_{abcd}$ is the Riemann curvature tensor)  for a given
spacetime is a sufficient condition to have spacetime singularities.
Therefore, we calculate the same for the gamma metric and get
\be
{\cal{K}} = \frac{16m^2\gamma^2 \ N}
               { r^{2\gamma^2+2\gamma+2} \lt(r-2m\rt)^{2\gamma^2-2\gamma+2}
                \Sigma^{3-2\gamma^2} },
\label{eq11}
\ee
where
\bea
N &=& m^2 \sin^2\theta \left\{ 
                         3m \gamma \lt(\gamma^2+1\rt) \lt(m-r\rt)
                     +\gamma^2 \lt(4m^2-6mr+3r^2\rt)+m^2\lt(\gamma^4+1\rt) 
                    \right\} \nn\\
   &+& 3 r \lt(\gamma m +m -r\rt)^2 \lt(r-2m\rt).
\label{eq12}
\eea
We study the nature of $r=2m$ spacetime singularity.
A spacetime singularity is called globally visible if there is a future
directed causal curve with one end ``on the singularity'' and the other end 
on the future null infinity. We investigate the polar as well as
the equatorial ``radial'' null geodesics in the gamma spacetime.
The Kretschmann invariant along these geodesics are, respectively,
\be
{\cal{K}}_{\lt(\theta=0\rt)} = \frac{48m^2\gamma^2\lt(m\gamma+m-r\rt)^2}
                              {r^{2\gamma+4} \lt(r-2m\rt)^{4-2\gamma}}
\label{eq13}
\ee
and
\be
{\cal{K}}_{\lt(\theta=\pi/2\rt)} 
= \frac{16m^2\gamma^2 \ S}
       {r^{2\gamma^2+2\gamma+2} \lt(r-2m\rt)^{2\gamma^2-2\gamma+2}
                                \lt(r-m\rt)^{6-4\gamma^2}},
\label{eq14}
\ee
where
\bea
S &=& m^4 \lt( \gamma^4+3\gamma^3+4\gamma^2+3\gamma+1 \rt)
   -3m^3r \lt( \gamma^3+4\gamma^2+5\gamma+2 \rt) \nn\\
   &+&3m^2r^2 \lt( 2\gamma^2+6\gamma+5 \rt) 
   -6mr^3 \lt( \gamma+2 \rt)
   +3r^4 .
\label{eq15}
\eea
${\cal{K}}_{\lt(\theta=0\rt)}$ diverges at $r=2m$ for  $\gamma<2$
($\gamma \neq 1$) only, whereas  ${\cal{K}}_{\lt(\theta=\pi/2\rt)}$ 
diverges at $r=2m$ for all values of $\gamma$ ($\gamma \neq 1$).
As the divergence of the Kretschmann invariant is a sufficient
(not the necessary) condition for a spacetime singularity, one concludes
that $r=2m$ is a curvature singularity in the gamma metric irrespective
of the value of $\theta$.

The null geodesics  are governed by equations
\be
\frac{dv^i}{dk} + \Gamma^i_{jk} v^j v^k = 0,
\label{eq16}
\ee
with
\be
g_{ij} v^i v^j = 0.
\label{eq17}
\ee
where $v^i \equiv \frac{dx^i}{dk}$ is the tangent vector to the
null geodesics ($k$ is the affine parameter).
\begin{flushleft}
{\underline{Case(i) \ \ Outgoing polar ``radial'' null geodesics}}
\end{flushleft}
The outgoing polar ``radial'' null geodesics in gamma spacetime are given by
\bea
v^t&=&E \lt(1-\frac{2m}{r}\rt)^{-\gamma}, \nn\\
v^r&=&E, \nn\\
v^{\th}&=&v^{\phi}=0,
\label{eq18}
\eea
where $E (E>0)$ is an integration constant. Thus, one has
\be
dt = \lt(1 - \frac{2m}{r}\rt)^{-\gamma} dr
\label{eq19}
\ee
We evaluate the integral 
$\lim_{\epsilon \rightarrow 0}  \int^R _{2m + \epsilon} 
\lt(1 - \frac{2m}{r}\rt)^{-\gamma} dr $, where R is finite.

For $\gamma > 1$,
\bea
\lim_{\epsilon \rightarrow 0}
 \int^R _{2m + \epsilon} \lt(1 - \frac{2m}{r}\rt)^{-\gamma} dr
 &>& \lim_{\epsilon \rightarrow 0}
\lt(2m +\epsilon\rt)^{\gamma}
 \int^R _{2m + \epsilon} \frac{dr}{\lt(r- 2m \rt)^{\gamma} } \nn\\
&=& \lim_{\epsilon \rightarrow 0}
 \frac{\lt(2m +\epsilon\rt)^{\gamma}}{\lt(\gamma - 1 \rt)}
 \lt[\epsilon^{1 -{\gamma}} - \lt(R -2m \rt)^{1 - {\gamma}}\rt] 
= \infty.
\label{eq20}
\eea
For $\gamma < 1$,
\bea
\lim_{\epsilon \rightarrow 0}
 \int^R _{2m + \epsilon} \lt(1 - \frac{2m}{r}\rt)^{-\gamma} dr
&<& \lim_{\epsilon \rightarrow 0}
R^{\gamma} \int^R _{2m + \epsilon} \frac{dr}{\lt(r- 2m \rt)^{\gamma} }\nn\\
&=& R^{\gamma} \frac{\lt(R-2m \rt)^{1 - \gamma}}{\lt(1 - \gamma\rt)},
\label{eq21}
\eea
which is  finite. Thus, with respect to the polar ``radial''
null geodesics, the $r=2m$ singularity is not (even locally) naked for 
$\gamma>1$, whereas it is globally visible for $\gamma<1$. 
\begin{flushleft}
{\underline{ Case(ii)\ \  Outgoing equatorial ``radial'' null geodesics}}
\end{flushleft}

The outgoing equatorial ``radial'' null geodesics in the spacetime are  given
 by
\bea
v^t&=&E \lt(1-\frac{2m}{r}\rt)^{-\gamma}, \nn\\
v^r&=&E\lt(\frac{r^2-2mr+m^2}{r^2-2mr}\rt)^{\frac{\gamma^2-1}{2}}, \nn\\
v^{\th}&=&v^{\phi}=0,
\label{eq22}
\eea
Thus, one has
\bea
dt &=& \lt(1 - \frac{2m}{r}\rt)^{-\gamma} 
\frac{\lt(r^2 - 2mr\rt)^\frac{\gamma^2-1}{2} }
{\lt(r - m \rt)^{\gamma^2 -1}}\  dr \nn\\
&=&\lt[r^\frac{2\gamma - \gamma^2 +1}{2} \lt(1 - 
                \frac{m}{r}\rt)^{1 - \gamma^2}\rt] 
\frac{dr}{ \lt(r -2m \rt)^{\frac{2 \gamma - \gamma^2 +1}{2}}         }
\label{eq23}
\eea
For finite  $R$,

$
\lim_{\epsilon \rightarrow 0} \ \int^R _{2m + \epsilon}
\lt[r^\frac{2\gamma  - \gamma^2 +1}{2} \lt(1 - 
                \frac{m}{r}\rt)^{1 - \gamma^2}\rt] 
\frac{dr}{ \lt(r -2m \rt)^{\frac{2 \gamma - \gamma^2 +1}{2}}         }
$

is clearly positive and finite for all values of $\gamma$
($\gamma \neq 1$). Therefore, with respect to the equatorial ``radial''
null geodesics, $r=2m$ singularity is globally naked for all values of
$\gamma$. Thus, the $r=2m$ spacetime singularity in the axially symmetric
gamma spacetime has ``directional nakedness'' for $\gamma>1$,
i.e., it is globally visible along $\theta =\pi/2$, whereas it
is not (even locally) naked along $\theta=0$. However, for $\gamma<1$,
$r=2m$ singularity is globally naked along $\theta=0$ as well as
$\theta=\pi/2$. It is of interest to investigate whether or not this
singularity ($r=2m$ for $\gamma>1$) is naked along other directions.
As the ``directional naked'' singularities can exist only in aspherical
spacetimes, it is of interest to study if it is a generic feature
of such spacetimes (of course,  with some restrictions on the spacetime
parameters, e.g., $\gamma>1$ in the gamma metric and $m^2<a^2$
in the Kerr metric). It remains to be investigated whether or not a 
``directional naked'' singularity  occurs in the collapse from a reasonable
nonsingular initial  data. The detailed studies of these would enrich our
knowledge of singularities in general  relativity and this could
have implications for astrophysics.

\begin{flushleft}
{\large Acknowledgements}
\end{flushleft}
Thanks are due to Kerri (R.P.A.C.)  Newman for helpful correspondence
throughout preparation of this work.


\newpage
\begin{thebibliography}{99}
\setlength{\parskip}{0.32ex}

\bibitem{Pen69} R. Penrose, Riv. del. Nuovo Cim. {\bf 1}, 252 (1969).
\bibitem{Pen78} R. Penrose, in {\em Theoretical Principles in
    Astrophysics and Relativity}, edited by N. R. Lebovitz {\em et
    al} (The University of Chicago press, Chicago, 1978) p.~217.
\bibitem{Haw79} S. W. Hawking, Gen. Relativ. Gravit. {\bf 10}, 1047  (1979).
\bibitem{PenSeiNew} R. Penrose, in {\em General Relativity - an Einstein
    Centenary Survey}, edited by S. W. Hawking and W. Israel (Cambridge
    University  Press, Cambridge, England, 1979), p.~581;
    R. P. A. C. Newman, Class. Quantum Grav.{\bf 3}, 527 (1986).
\bibitem{Pen72} R. Penrose, Nature {\bf 236} April 21, 377 (1972).
\bibitem{Pen73} R. Penrose, Ann. N. Y. Acad. Sci. {\bf 224}, 125
    (1973).
\bibitem{HawWalNabCla} S. W. Hawking and G. F. R. Ellis,{\em
     The Large Scale Structure  of Space-Time} (Cambridge University Press,
         Cambridge, 1973);
     R. M. Wald, {\em General Relativity} (The University of Chicago 
        Press, Chicago, 1984);
    G. L. Naber, {\em Spacetime and Singularities an Introduction}
       (Cambridge University Press, Cambridge, 1988);
    C. J. S. Clarke, {\em The Analysis of Space-Time Singularities}
       (Cambridge University Press, Cambridge, 1993).
\bibitem{TipNew} 
    F. J. Tipler, C. J. S. Clarke and G. F. R. Ellis, in  {\em General
          Relativity and Gravitation}, edited by A. Held  (Plenum, NY, 1980)
           p.~97;
    R. P. A. C. Newman, Gen. Relativ. Gravit. {\bf 15}, 641 (1983);
                        Gen. Relativ. Gravit. {\bf 16}, 1163 (1984);
                        Gen. Relativ. Gravit. {\bf 16}, 1177 (1984);
                        Proc. R. Soc. Lond. {\bf A443}, 473 (1993);
    E. Malec,{\em Global solutions of a free boundary problem for 
          selfgravitating scalar fields}, gr-qc/9506005;  
          {\em Selfgravitating nonlinear scalar fields}, Commun. Math. Phys.,
          1996 (submitted).
\bibitem{NS}
    P. Yodzis, H. J. Seifert, and H. M. Hagen, Commun. Math.
           Phys. {\bf 34}, 135 (1973);    
    B. Steinm\"{u}ller, A. R. King, and J. P. Lasota, Phys. Lett.
          {\bf 51A}, 191 (1975);
    Y. Kuroda, Prog. Theor. Phys. {\bf 72}, 63 (1984);
    D. Christodoulou, Commun. Math. Phys. {\bf 93}, 171 (1984);
    A. Papapetrou, in {\em A Random Walk in General Relativity},
            edited by N. Dadhich {\em et al} (Wiley Eastern, New Delhi,
            1985) p.~184;
    A. Ori and T. Piran,  Phys. Rev. {\bf D42}, 1068 (1990);
    S. L. Shapiro and S. A. Teukolsky, Phys. Rev. Lett. {\bf 66}, 
          994 (1991);
    K. Lake, Phys. Rev. Lett. {\bf 68}, 3129 (1992);
    J. P. S. Lemos, Phys. Rev. Lett. {\bf 68}, 1447 (1992);
    M. W. Choptuik, Phys. Rev. Lett. {\bf 70}, 9 (1993);
    D. Christdoulou, Ann. Math. {\bf 140}, 607 (1994);
    T. P. Singh and P. S. Joshi, Class. Quantum Grav. {\bf 13}, 559
           (1996). 
\bibitem{Jos93} P. S. Joshi, {\em  Global Aspects in  Gravitation and 
                Cosmology}, (Clarendon Press, Oxford 1993).
\bibitem{NSP} 
    K. S. Virbhadra, S. Jhingan, and P. S. Joshi, {\em  Nature of singularity
           in Einstein-massless scalar theory}, gr-qc/9512030; 
\bibitem{UnnAnt}
    C. S. Unnikrishnan, Phys. Rev. {\bf D53}, R580 (1996);
    H. M. Antia, Phys. Rev. {\bf D53}, 3472 (1996).
\bibitem{Ste}
    B. W. Stewart, D. Papadopoulos, L. Witten, R. Berezdivin,
     and L. Herrera,  Gen. Relativ. Gravit. {\bf 14}, 97 (1982).
\bibitem{Wei}
    S.~Weinberg, {\em Gravitation and Cosmology: Principles and  Applications
       of General Theory of Relativity} (John Wiley and Sons, Inc.) p.~165.
\bibitem{ACV}
    J. M. Aguirregabiria, A. Chamorro, and K. S. Virbhadra, {\em
       Energy and angular momentum of charged rotating black holes},
       Gen. Relativ. Gravit., 1996 (to appear).



\end{thebibliography}
\end{document}